\documentclass[runningheads]{llncs}
\usepackage{graphicx}
\usepackage{algorithm}
\usepackage{algpseudocode}
\usepackage{comment}
\usepackage{amsmath}
\usepackage{subfig}

\algnewcommand\algorithmicforeach{\textbf{for each}}
\algdef{S}[FOR]{ForEach}[1]{\algorithmicforeach\ #1\ \algorithmicdo}

\begin{document}
\title{Run-time Norms Synthesis in Multi-Objective Multi-Agent Systems}
%
%
\author{Maha Riad\and
Fatemeh Golpayegani}
\authorrunning{M. Riad and F. Golpayegani}
%
\institute{School of Computer Science, University College Dublin, Ireland \email{maha.riad@ucdconnect.ie} \\ \email{fatemeh.golpayegani@ucd.ie}}

\maketitle              
\begin{abstract}
Norms represent behavioural aspects that are encouraged by a social group of agents or the majority of agents in a system. Normative systems enable coordinating synthesised norms of heterogeneous agents in complex multi-agent systems autonomously. In real applications, agents have multiple objectives that may contradict each other or contradict the synthesised norms. Therefore, agents need a mechanism to understand the impact of a suggested norm on their objectives and decide whether or not to adopt it. To address these challenges, a utility-based norm synthesis (UNS) model is proposed which allows the agents to coordinate their behaviour while achieving their conflicting objectives. UNS proposes a utility-based case-based reasoning technique, using case-based reasoning for run-time norm synthesising in a centralised approach, and a utility function derived from the objectives of the system and its operating agents  to decide  whether or not to adopt a norm.
The model is evaluated using a two intersecting roads scenario and the results show its efficacy to optimise multiple objectives while adopting synthesised norms.

\end{abstract}
\section{Introduction}
 Multi-agent systems (MAS) are complex systems consisting of agents which are autonomous entities with their own objectives, and can act dynamically. Agents' objectives can be represented by tasks they want to achieve, these tasks can be unintentionally supportive to other agents’ objectives or contradict them \cite{golpayegani2019using}. Aside from the ability of the agents to have multiple objectives, agents may have heterogeneous types, in which each type has its own characteristics, preferences or category \cite{ghanadbashiusing}. Moreover, agents can operate in open system settings where they can move freely from one system to another. MAS is applied in many real world applications such as traffic systems \cite{bailey2019comasig} \cite{ghanadbashiusing}, computer networks \cite{dorri2018multi}, smart energy grids \cite{golpayegani2016multi} and the internet of things systems \cite{singh2017internet}. However, in such systems, it is not only crucial to model the heterogeneity, openness and autonomy of the agents, but also it is essential to consider the agents' behaviour coordination.  

Norms are behaviour guidelines imposed by a society or social group to regulate agents’ actions. For example, in a traffic system, one norm is to slow down when seeing a senior driver because he might be more cautious than other drivers and drive slowly. Another example is represented in the norm leaving the right (fast) lane empty when there is an ambulance. Accordingly, norms representation helps agents to achieve their objectives in an acceptable manner within their social groups without compromising their autonomy. This would facilitate group decision making, cooperation, and coordination between agents \cite{oliveira2016analysis}. 

Multi-agent systems that encapsulate norms concepts such as prohibitions, obligations and permissions are called Normative Multi-Agent Systems (NorMAS) \cite{dastani2018normative}. NorMAS rely on norms for regulating the behaviour of agents while reserving their autonomy property \cite{dastani2018normative}. Norms have dynamic nature, and so each norm's life cycle begins with norm synthesis (which relies on creating and composing a set of norms \cite{morales2013automated}) and ends with  norm disappearance \cite{frantz2018modeling}. 

The process of norm synthesis is challenging as it may produce unmatchable norms or norms that contradict agents' objectives. For example, it might be a norm that a driver drive in an average or a slow speed when having a child on board. However, the same car might drive too fast when the child has an emergency. So, in this case two different unmatchable norms appear: (i) a car drive slowly if a child is on board, and (ii) a car drive too fast in case of having an emergency case. As an example of norms contradicting agents' and systems' objectives, consider the following case in traffic context, a norm might be to give priority to an emergency car, such as ambulance and police car, although this might contradict the agent's objectives such as  arriving at the minimum time or having the minimum number of stops.

To address the challenges of handling unmatchable norms, and coordinating norms and objectives, we present a utility-based norm synthesis (UNS) model.  In UNS, a utility-based case-based reasoning technique is proposed to facilitate the coordination of norms and objectives of agents and the system. UNS uses the case-based reasoning algorithm to synthesise norms. The utility function determines the necessity of norms adoption and elicits the suitable norm when there are unmatchable norms. Two norms are called unmatchable norms, when only one of them should be applied at the same and context. For example, consider norm $n_a$, which suggests to stop if there is a car on the left side of a junction, and norm $n_b$, recommending to stop if there is a car on the right side of a junction. Although by applying these two norms a collision would be avoided, a deadlock situation will be created as well.  The utility function is constructed based on the objectives of the system to ensure that they are considered in the process of norms reasoning. UNS is evaluated using a simulated traffic scenario in SUMO and results show the system's capability of synthesising and reasoning norms in run-time while reaching the system’s objectives.
\section{Related Work}
Synthesising norms are more challenging in open and run-time NorMAS.  In open systems, the challenge is to transfer norms to new agents entering the system and make use of the norms adopted by other agents before leaving the system.  Mahmoud \emph{et al.} \cite{mahmoud2016development} address this challenge by proposing a potential norms detection technique (PNDT) for norms detection by visitor agents in open MAS. They implemented an imitating mechanism which is triggered if visitor agents, who are monitoring norms of host agents, discovered in-compliance in their norms with the other host agents. However, PNDT technique used a fixed set of norms, which are commonly practised by the domain, ignoring the dynamic nature of norms. 

 In run-time NorMAS, it is challenging to define a dynamic set of norms and initiate it. Moreover, real run-time applications would not only demand synthesising new norms but would also require handling the whole norms life-cycle including norms refinement and disappearance. One of the efforts directed towards run-time norms revision was carried out in \cite{dell2018runtime} in which a supervision mechanism for run-time norms revision was proposed, addressing the challenge of norms modification when the weather changes or accidents happen. However, the norms revision mechanism is developed using a primary defined pool of norms and situations. In the revision process, the model just substitutes the norms depending on the situation; limiting the norms to a static set of norms. Accordingly, the dynamism is in altering the chosen norms set depending on an optimisation mechanism for the system objectives and does not handle the norms changes and evolution. In \cite{edenhofer2016bottom}, Edenhofer \emph{et al.}. present a mechanism for dynamic online norm adaption in a heterogeneous distributed multi-agent system for handling colluding attacks from agents with bad behaviour. The agents interact together and build a trust metric to represent the reputation of the other agents. The main focus of this paper is identifying the bad agents and showing that using norms improves the system's robustness. Although this work is based on an open, heterogeneous and distributed environment, the research does not identify how norms can be revised and updated in this context. 
 
 IRON machine was developed by Morales \emph{et al.} and presented in \cite{morales2013automated}. The main aim of IRON is to synthesise norms on-line using an effective mechanism that not only synthesises norms in run-time but also revises these synthesised norms according to their effectiveness and necessity and further dismisses the inefficient norms. IRON simulates multi-agent systems, in which norms are synthesised for coordinating the behaviour of agents, and handles conflicting situations that can occur, such as collisions of vehicles in a traffic scenario. As presented in \cite{morales2013automated}, IRON is capable of run-time norm synthesising and addressing the issues of using static norms, however, the idea of coordinating norms and objectives is not addressed. 
 
 

\subsection{Intelligent Robust On-line Norm synthesis machine (IRON)}

IRON machine is composed of a central unit that is responsible for detection of conflicts, synthesises of new norms to avoid conflicts, evaluation of the synthesised norms, refinement of norms, and announcement of the norm set to the agents. To simplify the illustration of the responsibilities of IRON, we will use a traffic junction example with two orthogonal roads scenario. The vehicles  represent the agents, each  occupying  a single cell and moving in a specific direction per time-step. 
\begin{itemize}
    \item In {\bfseries conflicts detection},
conflicts are detected when a collision occurs between two or more vehicles. The occurrence of a collision will trigger IRON to synthesise new norm to avoid future collisions of similar cases.
As for {\bfseries norms synthesising},
norms are created based on a case-based reasoning algorithm. In the algorithm, the conflicting situation at time $t$ is compared to the conflicting vehicles’ context at time $t-1$. Then a norm is created using the conflicting views as a precondition for applying the norm and prohibiting the ‘Go’ action in this context. The synthesised norm is then added to a norm set and communicated to the agents(vehicles) of the system. For example, in Fig. \ref{fig2}, if vehicle $A$  and $B$ collided at the intersection (grey cell) then the context and action of $A$ or $B$ is chosen randomly by the system to create a new norm. If $A$ is chosen the new norm will be $n=if(left(<),front(-),right(<))->proh('Go')$. The left() attribute in the precondition of the norm stores the direction of the left neighbour vehicle of vehicle $A$. While the right() attribute stores the direction of the right vehicle to vehicle $A$, which is in this case vehicle $B$.
\item {\bfseries Norms Evaluation}
is carried out by measuring necessity and effectiveness of a norm and comparing it to a threshold. Necessity is measured according to the ratio of harmful violated norms, which are norms that resulted in conflicts when violated, compared to the total number of violated norms. The norm's necessity reward $NNR$ is calculated by:

\begin{equation}
 NNR=\frac{m_{V_C}(n)\times w_{V_C}}{m_{V_C}(n)\times w_{V_C}+m_{V_{\bar{C}}}(n)\times w_{V_{\bar{C}}}}
\end{equation}
\\
$
m_{V_C}(n)$: Number of violations which led to conflicts 

$w_{V_C}$: Weight that measure the importance of harmful applications

$m_{V_{\bar{C}}}(n)$: Number of violations which did not led to conflicts

$w_{V_{\bar{C}}}$: Weight that measure the importance of harmless applications
\\

The effectiveness of norms is measured based on the extent to which the norm is successful (i.e. which resulted in the minimum number of conflicts). The norm's effectiveness reward $NER$ is calculated by:

\begin{equation}
 NER=\frac{m_{A_C}(n)\times w_{A_C}}{m_{A_C}(n)\times w_{A_C}+m_{A_{\bar{C}}}(n)\times w_{A_{\bar{C}}}}
\end{equation}
\\
$
m_{A_C}(n)$: Number of applied norms which led to conflicts

$w_{A_C}$: Weight that measure the importance of unsuccessful applications

$m_{A_{\bar{C}}}(n)$: Number of applied norms which did not led to conflicts

$w_{A_{\bar{C}}}$: Weight that measure the importance of successful applications
 \\

\item {\bfseries Norms refinement}
is carried out by generalisation or specialisation of norms. Norms are mapped in a connected graph that expresses the relationship between the norms. The graph shows the child and parent norms that are linked. Norms generalisation is applied when two or more norms have acceptable necessity and effectiveness results compared to the threshold for time-interval $T$. Specialisation or deactivation of norms is conducted when the effectiveness and necessity of the norm or its children have been below threshold for time-interval $T$.
\item {\bfseries Norms communication}
is the final step, where the norms are communicated to the agents.

\end{itemize}
The main flow of activities that are carried out in the traffic scenario with the traffic junctions of two orthogonal roads (similar to Fig. \ref{fig2}), is as follows. Agent movements take place per time-step, however, prior to this movement, the norms set is checked for norms with matching precondition. If a norm was found, it will be applied directly, assuming there are no bad-behaving agents who violate the norms. If no norm was found the movement is undertaken.
When a new collision is detected, a random agent/vehicle is chosen and its context (local view) is added as a precondition of a new norm that prohibits the ‘GO’ action. Then this norm will be added to the norms set (initially empty). In addition, norms evaluation and refinement are carried out per time-step, in which all the views at time-step $t$ are revised to determine the set of applicable norms for each of the views. Afterwards, the applicable norms are divided into four subsequent sets: (i) applied norms that led to conflicts (ii) applied norms that did not lead to conflicts (iii) violated norms that led to conflicts (iv) violated norms that did not lead to conflicts. Then set (i) and (ii) are used to calculate the effectiveness of each of the norms, while set (iii) and set (iv) represent the main inputs for the necessity calculation. Finally, norms refinement is conducted. 

\section{Problem Statement}
Let us consider a norm-aware multi-objective multi-agent system that is composed of a finite set of mobile agents as $Ag=\{ag_1,ag_2,...,ag_n\}$. Each agent $ag_i$ has a type $t_{ag_i}$, set of properties $P_{ag_{i}}$, set of objectives $O_{ag_{i}}$ and set of adopted norms $N_{ag_{i}}$. In addition, the system itself has its own set of objectives $O_s$ and set of norms $N_s$, where $O_{ag_{i}} \subseteq O_s$  and $N_{ag_{i}} \subseteq N_s$. 

The norms are created by a centralised unit in the system in the form of a pair $(\alpha,\theta(ac))$ and then messaged to the agents. $\alpha$ represents a precondition for triggering the norm applicability. This precondition reflects a specific context of the agent $c(ag_i)$, which is the local view of  agent $ag_i$ that defines its direct neighbours $Ng_{ag_{i}}=\{ag_1,ag_2,...,ag_k\}$ and their properties such as their moving direction in traffic scenario example. So, $c(ag_i)=\{P_{ag_{k}}:ag_k \subseteq Ng_{ag_{i}}\}$. $\theta$ symbolises a deontic operator (obligation, prohibition or permission) with a specific action $ac_{ag_i}$ of agent $ag_i$ which will apply the norm. For example, if an action is beneficial for an agent then it is obligated and if an action is harmful it is prohibited.

The central unit synthesises new norms after a conflicting state $c$ arises between agents and uses the synthesised norm in future similar cases to avoid conflicts. Conflicting state $c$ belongs to set of conflicts $C$, a conflict is considered detected when two agents or more carry out actions that result in a problem. The norm is synthesised by comparing the view at conflicting situation at timestep $t$,  $V_t$ to the view before the conflict has occurred $V_{t-1}$. The series of views that represent different situations at each timestep are added in a ViewTransition $V$ set (i.e. $V_t$ $\in$ $V$ and $V_{t-1}$ $\in$ $V$).

In such a system, there are three main problems to be tackled. First, the process of synthesising norms, which should ensure fairness (i.e. created norms cannot be biased towards specific agents' situation). For example, if there is a norm  created to coordinate the behaviour of two vehicles $ag_1$ and $ag_2$ in an intersection, this norm cannot always give priority to the vehicles on the right, because this will make the vehicles in the left lane always delayed. Second, when there are more than one norm, often unmatchable ones, to coordinate the behaviour in a situation, 
only one should be applied to avoid a deadlock situation. For example, in a scenario of vehicles crossing a junction, if there were two norms created: $n_1$ for stopping if there was a vehicle on the right, and $\acute{n}_1$ for stopping if there is a vehicle on the left, a decision should be made to apply one of these unmatchable norms only. Third, agents' norms $N_{{ag}_i}$ and agent's objectives $O_{{ag}_i}$ should be coordinated to ensure that applying norms do not contradict their objectives.

\section{UNS: Utility-Based Norm Synthesis Model}
\begin{figure}
\includegraphics[width=\textwidth]{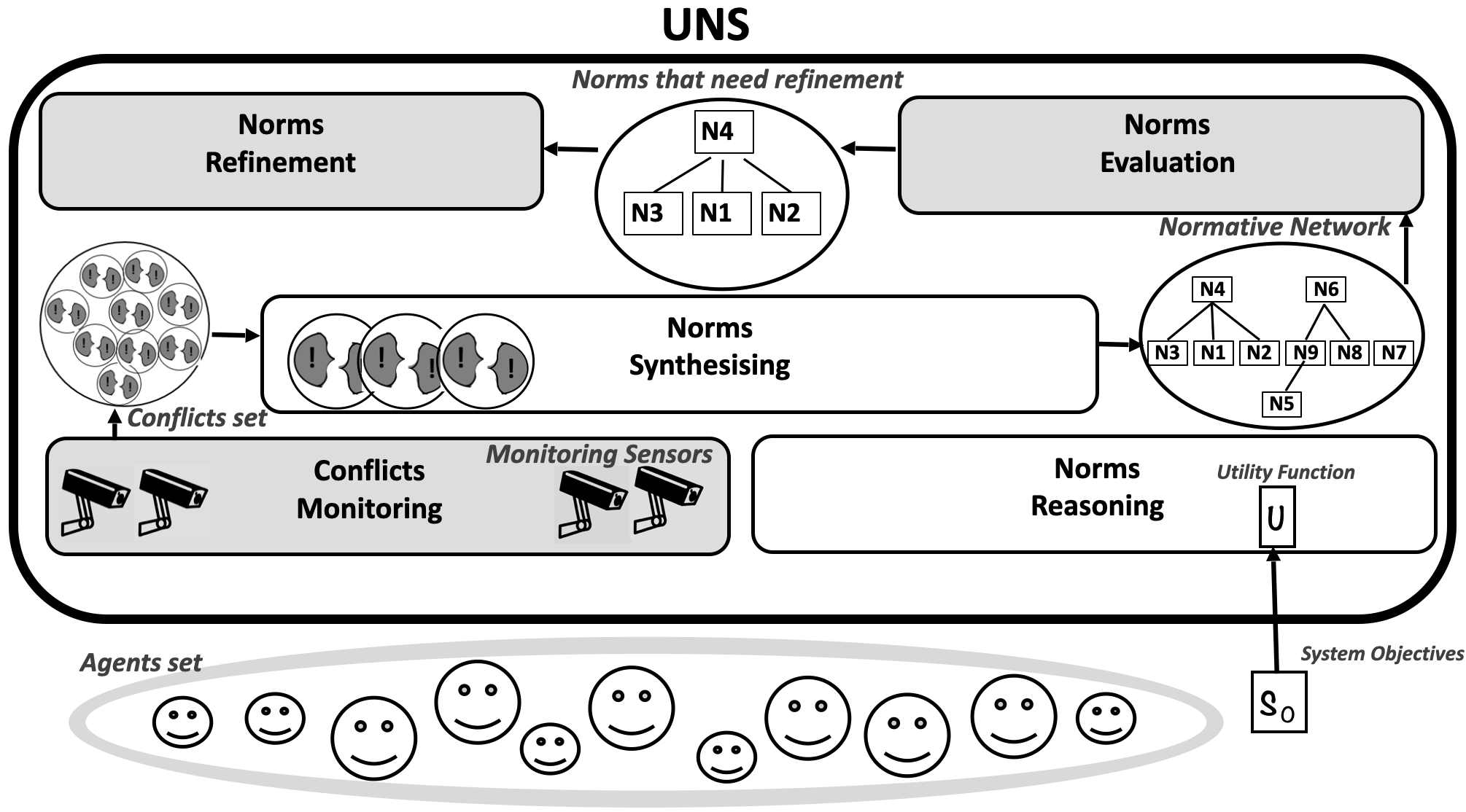}
\caption{Utility-Based Norm Synthesis Model Architecture
(components coloured in grey are inherited from IRON)} \label{fig1}
\end{figure}

UNS is a utility-based norm synthesis mechanism implemented in a normative, open, run-time, multi-objective, multi-agent system. UNS aims at reaching three main goals. First, to synthesise norms while supporting fairness during norm creation. Second, to handle unmatchable synthesised norms. Finally, to coordinate the objectives of agents with the synthesised norms. Fig. \ref{fig1} shows the architecture of UNS. It shows the five main responsibilities of UNS that takes place per time-step at run-time: conflicts detection, norms synthesising, norms reasoning, norms evaluation and refinement.

Conflicts detection, norms evaluation and refinement are inherited from IRON and integrated in UNS. The details of the steps carried out by UNS are as follows:  
\subsection{Conflicts Detection}
At each time-step $t$ as agents take actions, a set of monitors (e.g. traffic cameras) $M=\{m_1,m_2,...,m_n\}$ monitor these actions to detect any conflicts. 
A conflict $c$ is detected when more than one agent actions contradict at the same view $v_i$, where $v_i\in V$ . For example, in a traffic system, if vehicles standing before a junction in opposite directions decided to move (do a 'Go action') towards the same position, a collision will occur and so a conflict will arise. To detect conflicts, views $V$ are sent as a parameter to the $ConflictDection$ function (see Algorithm 1, line 6). A conflict object definition is composed of responsible agents $Ag^r \subseteq Ag$ , context of these agents $ag^r(ac)$ (which is the local views of each of these agents), and the views transition of a state $s$ between time-step $t$ and $t-1$,  $(v^i_{s_{t-1}},v^i_{s_{t}})$. 
\subsection{Norms Synthesising}
Case-based reasoning technique is used for norm synthesising. When a new conflict arises, a new case is created and then compared to similar cases and the best solution is chosen accordingly. In case that no similar solution is found a new random solution is created for this case and added to the set of cases. In this manner, after conflicts are detected, UNS carries out the norms synthesising steps for each of these conflicts (see Algorithm 1,  line 7 to line 16). All the agents responsible for the conflict are retrieved in $Ag^r$ (e.g. all the vehicles that collided in the same intersection are considered as responsible agents). For each of these agents' context at $t-1$ if an applicable norm was not found (applicable norms are norms that have the same context as a pre-condition of the norm and the same agent action prohibited in the norm), a new norm creation process takes place (line 12). A new norm is composed of  agents' context $ag_i(c)$ and prohibited action $\theta ag_i(ac)$. Getting the context of the agent at the previous time-step as a precondition of a norm and prohibiting the action that  resulted in a conflict avoids future conflicts that might rise in similar situations. After the norm is created it is added to the system's norms set $\Omega$ (line 13).

{\bfseries UNS Supporting Fairness:}
In IRON norms synthesising was carried out by creating norms as a solution for only one randomly chosen agent from the agents involved in a conflict. However, in UNS we have proposed a norm synthesising process, which considers all the contexts of the agents involved in a conflict. For example, in IRON if two vehicles had a conflict in an intersection, the norm will be created based on prohibiting a Go action of only one of the two vehicles. Although this will decrease the probability of creating unmatchable norms, 
it will not ensure fairness as one side will always have priority over the other side. 


\begin{algorithm}
\caption{UNS Strategy}
\begin{algorithmic} [1]
\ForEach {$t$} \\
    \hspace*{\algorithmicindent} \textbf{Input:} $\Omega$, $V$, $Ag(ac)$ \\
    \hspace*{\algorithmicindent} \textbf{Output: $\Omega$} \\
    
    \textbf{/*Conflicts Detection*/} \\
    //Inherited from IRON
    \State $conflicts\gets ConflictDetection(V)$ \\
    \textbf{/*Norms Synthesis*/} 
    \ForEach {$c \in conflicts$} 
        \State $Ag^r\gets AgentsInConflictContexts(c)$ 
        \ForEach {$ag(c) \in Ag^r$} 
            \If{$hasApplicableNorm(ag(c))==false$}
            \State $n\gets CreateNorm(ag(c),\theta (ag(ac)))$
            \State  $\Omega \gets \Omega \cup{n}$
            \EndIf
        \EndFor 
    \EndFor \\
    \textbf{/*Norms Reasoning*/}
    \ForEach {$V_{t+1} \in V$} 
    \State $N_a\gets GetApplicableNorms(V)$ 
    \If {$N_a.size > 1$}
    \State $Utilities[] \gets null$     //comment:$Utilities[k]=(ag_i,n,U_i)$
    \ForEach {$n \in N_a$} 
    \State $Utilities.add(calculateUtility(n),n)$
    \EndFor 
    \State $ag_i \gets getAgent(max(Utilities))$
    \State $Ag.applyApplicableNorm()$
    \EndIf
    \EndFor \\
    \textbf{/*Norms Evaluation*/}\\
    //Inherited from IRON \\
    \textbf{/*Norms Refinement*/} \\
    //Inherited from IRON
\EndFor
\end{algorithmic}
\end{algorithm}

\subsection{Norms Reasoning}
 The norm reasoning process  must meet systems' objectives and  handle unmatchable norms simultaneously. This is reached through defining a utility function $U$ that is constructed based on the system's objectives $O_s$ and is used during the norm selection. 

{\bfseries Utility Function Construction:}
In this paper, the utility function is constructed  by adding the objectives with a maximisation function and subtracting the objectives with a minimisation function. For example, if  $S_o$ include two objectives $S_o=\{so_1, so_2\}$ and $so_1$ is to minimise all vehicles' average waiting time and $so_2$ is to minimise the average waiting time of emergency vehicles specifically, then the system utility function $U$ will be defined as: 
\begin{equation}
U= -so_1-so_2= -1*(so_1+so_2)
\end{equation}

\begin{figure}
\centering
\includegraphics[width=0.5\textwidth]{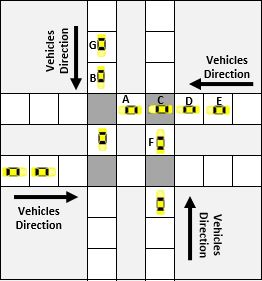}
\caption[width=0.5\textwidth]{A traffic junction composed of two orthogonal roads} \label{fig2}
\end{figure}

{\bfseries Accumulated Utility Calculation:}
 At each time-step before the agents start moving (taking actions) UNS determines the set of applicable norms $N_a$ in each view $V_t$ ( see  Algorithm 1, line 19). If more than one norm is applicable for the same view $V_t$, then UNS carries out the steps in Algorithm 1 (from line 21 to line 26) to choose the norm with the highest utility and dismisses the rest of the norms. For example, if we have a traffic scenario as seen in Fig. \ref{fig2}, where vehicle $A$, $ag_1$, and vehicle $B$, $ag_2$, are willing to move to the same junction (coloured in grey) at time $t$, the stored view at time $t$ will be represent by ($V_t$). UNS will retrieve the set of applicable norms $A_n=\{n1,\acute{n}1\}$, where $n1$: is to stop if there is a vehicle on the right and $\acute{n}1$ is to stop if there is a vehicle on the left. $n1$ is suggested for vehicle $A$ and $\acute{n}1$ is suggested for vehicle $B$. If both vehicles apply the norms then none of them will move, which result in a  deadlock state. So, a decision must be made to choose only one of the two unmatchable norms. Accordingly, an empty array of struct is initialised (in line 21). The struct is composed of $ag_i$ (which is the responsible agent), $n$ (which is the applicable norm for this agent $ag_i$ situation), and $U_i$ (which is the calculated utility gained by the system if this norm $n$ is applied). For each of the applicable norms in $N_a$, the utility function is calculated (line 23). However, in our utility calculation strategy, we calculate an accumulated utility function, which does not only consider the utility gained by the agent applying the norm, but also considers all the agents that are indirectly affected by the norm adoption decision. For example, in Fig. \ref{fig2}, if vehicle $A$ will be the agent that will apply the norm and will stop if there is a vehicle on the right, it will force vehicles $C,D,E$ and $F$ to stop as well. While if vehicle $B$ decides to apply the norm $\acute{n}1$ and to stop, vehicle $G$ will be forced to stop as well. Based on this justification, to ensure gaining the actual maximum utility, UNS aggregates the utility of all the agents that are affected  directly and indirectly  with the norms adoption or dismissal. Then, the norm that gives the maximum utility is applied (lines 25 and 26), and the rest of the norms are dismissed.
\subsection{Norms Evaluation and Refinement}
The norm evaluation and refinement processes are inherited from IRON (illustrated in section 2.1). These processes are used to evaluate norms at run-time using efficiency and necessity equations (equation 1 and 2). If a norm's efficiency and necessity does not reach a certain threshold its refinement takes place and it is specialised or deactivated. Also, if a norm's efficiency and necessity exceeds a specified threshold it can be generalised.

\section{Empirical Evaluation}
In this section we show UNS capability to synthesise norms that support fairness, to handle unmatchable norms and to coordinate norms and objectives.
\subsection{Empirical Settings}
We simulate a traffic-based scenario, with a 19 x 19 grid as a road network with a junction of two orthogonal roads (see Fig. \ref{fig2}). Each road has two lanes; one for each direction. In Fig. \ref{fig2}, the cells coloured in grey show the four cells that represent the intersections. Vehicles are the agents and they have two main types: ordinary vehicles and high priority vehicles to represent heterogeneity.  The ratio of generating priority vehicles to ordinary vehicles is 12:100 respectively. Also, as it is an open MAS, vehicles can enter and leave the road network freely. Vehicles move per time-step aiming to reach their final destination which was randomly generated by the simulator at the beginning of the trip of the vehicle. In each time-step, the system randomly chooses the number of new vehicles (between 2 to 8 vehicles) to be emitted to start their trip. 
The system aims at avoiding conflicts (i.e., the collisions between vehicles) through the synthesised norms. Norms are defined as a pair that includes the agent context and the prohibited action. The agent context is the local view of the vehicle describing the direction of vehicles on its left, front, and right, which we call neighbouring vehicles. For example, in Fig. 2 the vehicles in the local context of vehicle $F$ are vehicles $A$,$C$ and $D$.
The action prohibited is a ‘Go action’ to avoid vehicle movement in future similar contexts. UNS synthesises norms and adds them to a norm set that is initially empty at the beginning of the simulation. When a stable normative system is reached the system converges.
The system has two main objectives,  minimising the average waiting time for all vehicles and minimising the total waiting time of priority vehicles. 
The utility function used in the norm reasoning is constructed based on the previous two objectives as follows:

\begin{equation}
-1 * ( \frac{X_{wt}+Y_{wt}}{X+Y} + Y_{wt}) 
\end{equation}

$X_{wt}$:Total waiting time of ordinary vehicles

$Y_{wt}$:Total waiting time of priority vehicles

$X$:Number of ordinary vehicles

$Y$:Number of priority vehicles

\subsection{Experiment Results}
To evaluate UNS's performance, three main scenarios are tested with the settings illustrated in the previous sub-section with varying violation rate of norms, which represents the ratio of agents obeying the adoption of the norms. UNS will be compared to IRON machine (explained in section 2.1). 
The average waiting time for all vehicles and the total waiting time of priority vehicles are reported to show the performance of UNS and IRON. 
 Moreover, the number of collisions is used to reflect the efficiency of the synthesised norms in avoiding conflicts. We present the moving average of  the results at every 50 time-steps obtained from 10 runs of simulation as plotted in Fig. \ref{fig3}, \ref{fig4}, and \ref{fig5}.

\subsection{Scenario A (Violation Rate 10\%)}
\begin{figure}%
    \centering
    \subfloat[\centering Average waiting time for all vehicle types]{{\includegraphics[width=5.66cm]{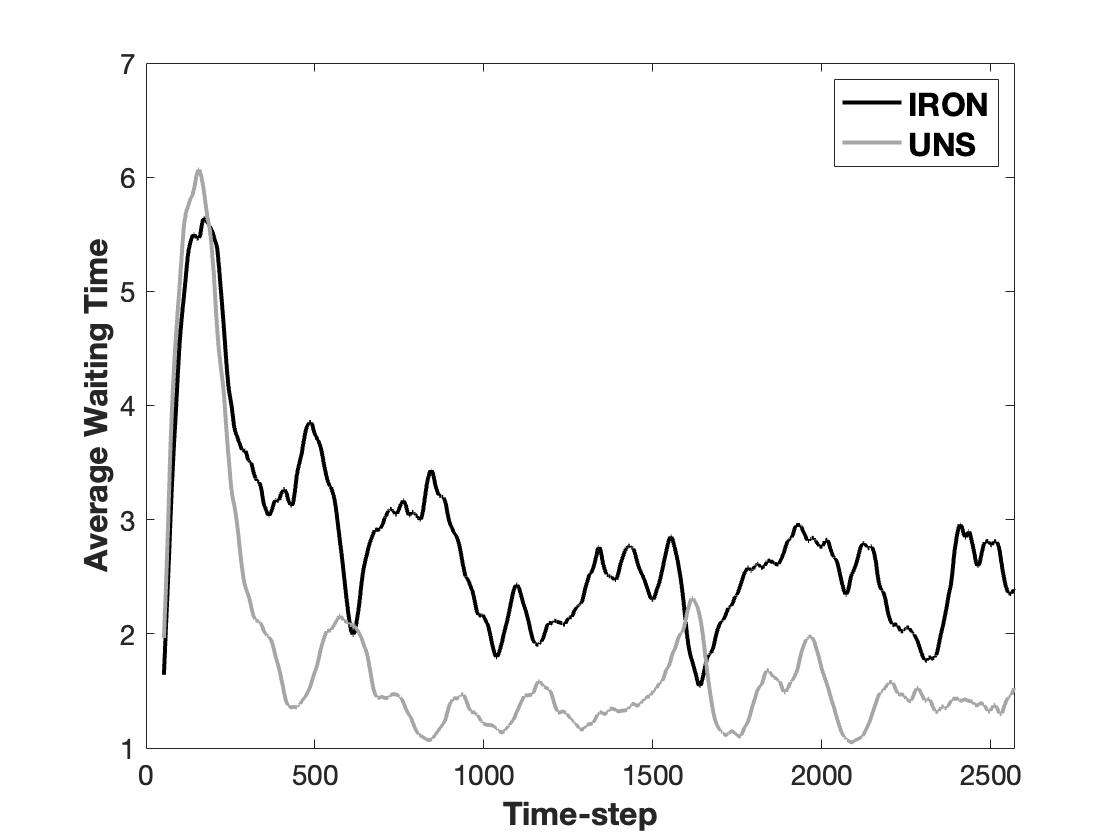}}}%
    \qquad
    \subfloat[\centering Total waiting time of priority vehicles]{{\includegraphics[width=5.66cm]{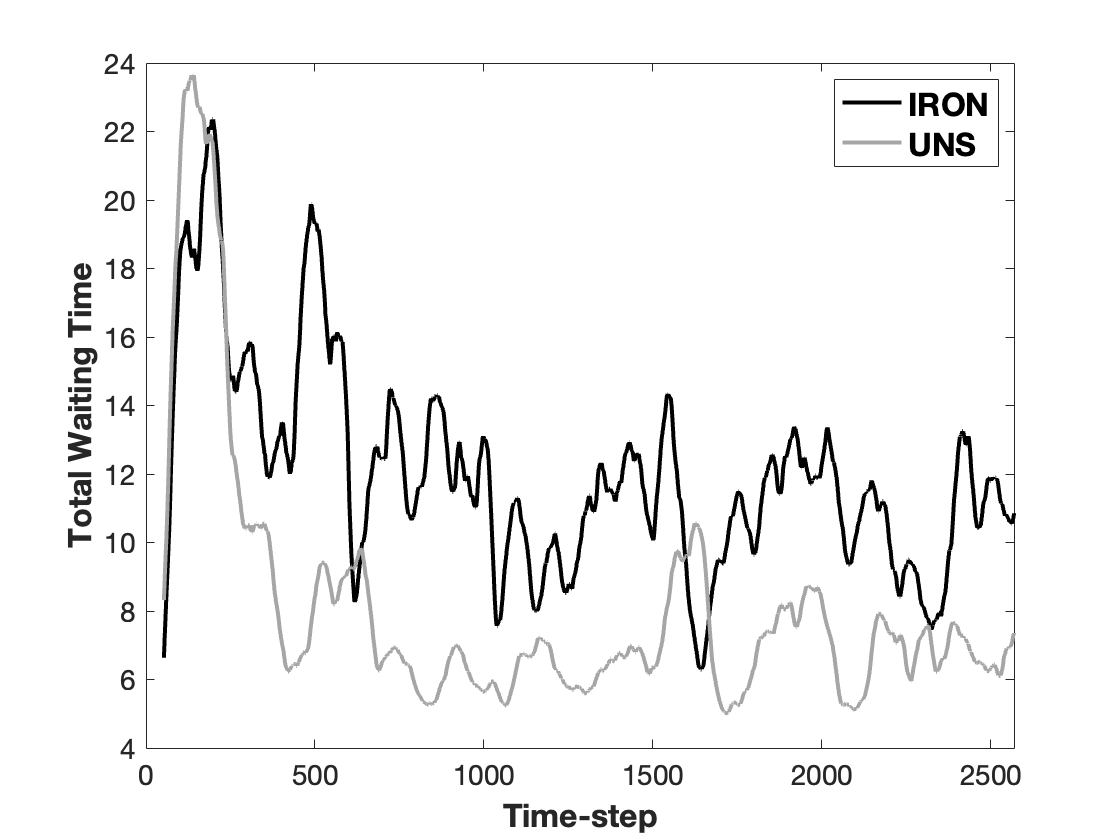}}}%
    \caption{Scenario A}%
    \label{fig3}%
\end{figure}

\begin{figure}%
    \centering
    \subfloat[\centering Average waiting time for all vehicle types ]{{\includegraphics[width=5.66cm]{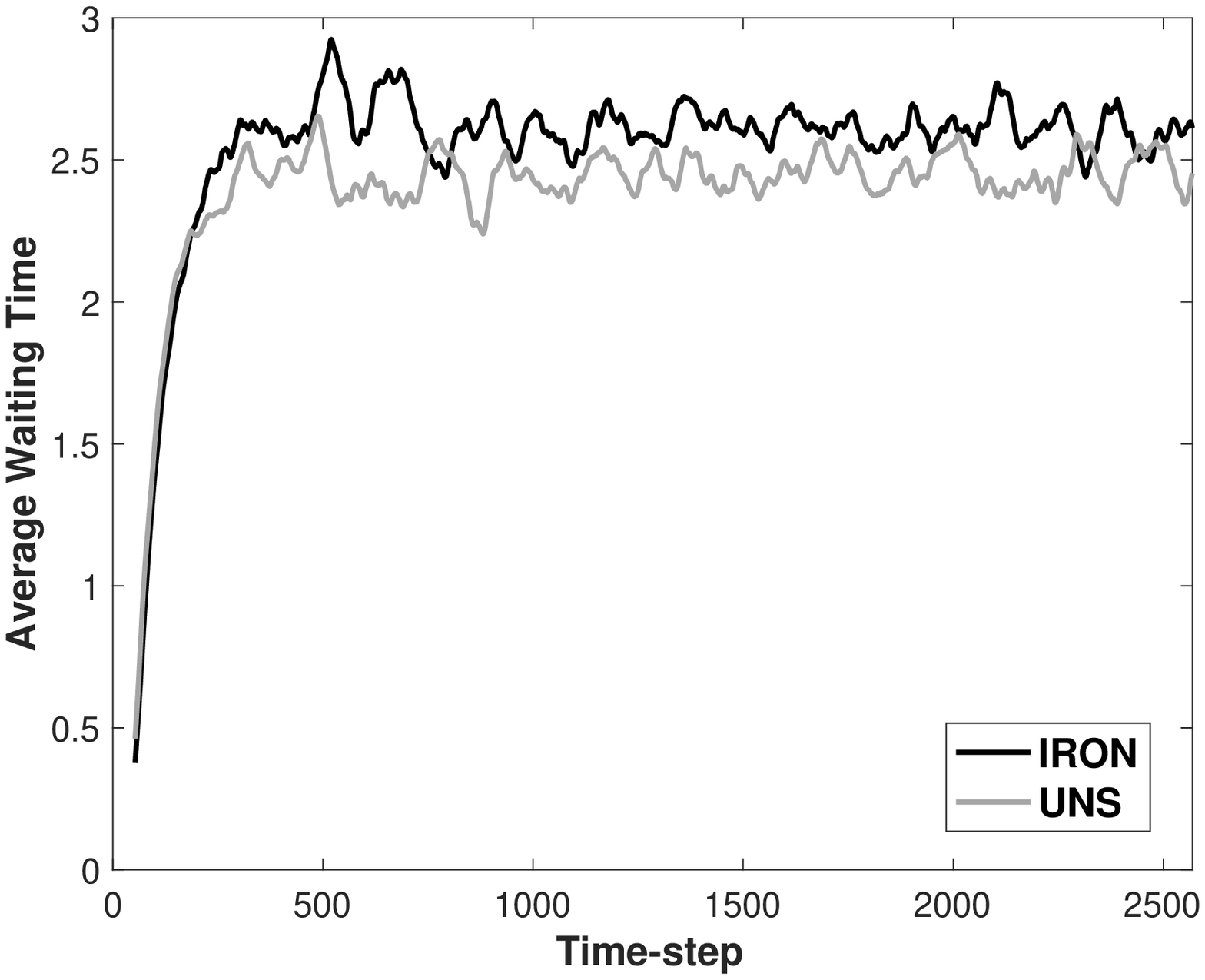}}}%
    \qquad
    \subfloat[\centering Total waiting time of priority vehicles]{{\includegraphics[width=5.66cm]{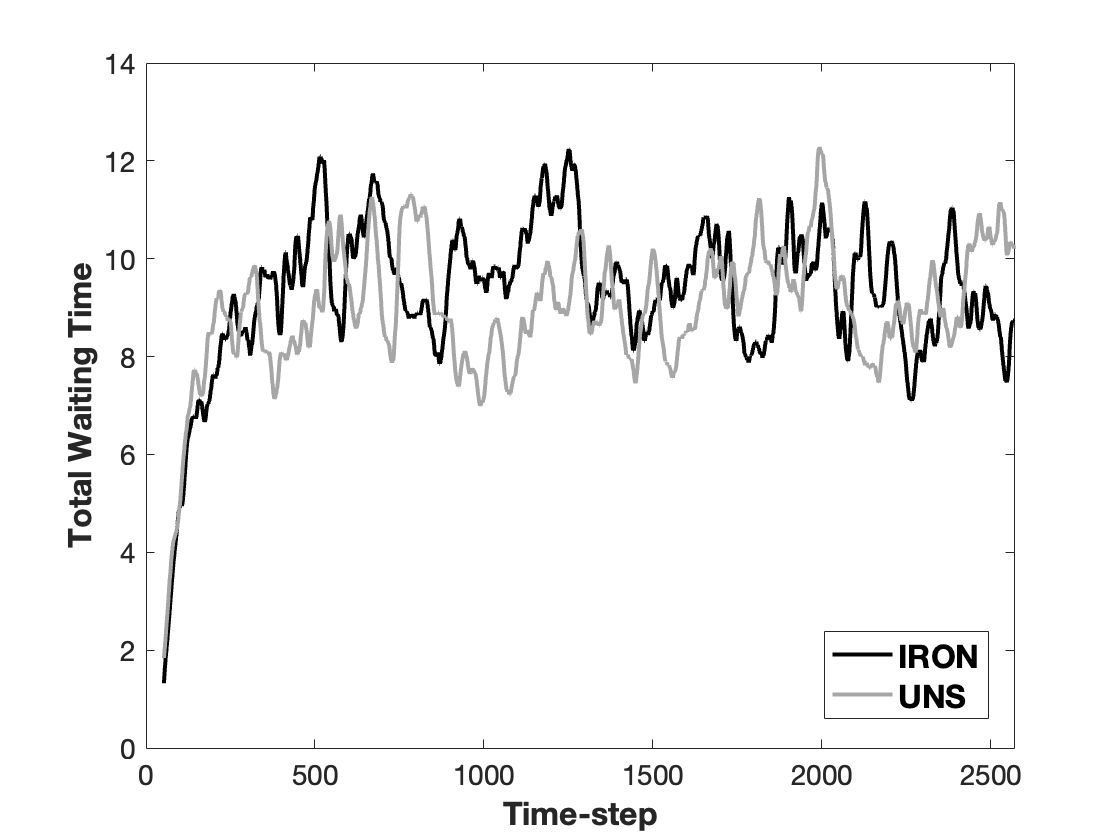}}}%
    \caption{Scenario B}%
    \label{fig4}%
\end{figure}

Fig. \ref{fig3}(a) shows the average waiting time of all vehicles in UNS compared to the average waiting time of all vehicles in IRON.  The average waiting time is decreased in UNS, particularly from time-step 322. Moreover, it can be noted that  from time-step 322 almost the average waiting time in UNS is constant  with an average value of 1.5 time-steps. As results show, UNS has minimised the average waiting time of the vehicles and so fulfilling the first objective of the system.

Fig. \ref{fig3}(b) shows the total time taken by priority vehicles per time-step in UNS compared to IRON. The average total waiting time of priority vehicles using UNS is 8.09 time-steps, while the average total waiting time of priority vehicles reached in IRON is 12 time-steps.  Moreover, Fig. \ref{fig3}(a) and (b) do not only emphasise how UNS can coordinate objectives and norms, but the noticed stability and uniformity of the results show the reliability of UNS which is necessary in  real-applications. 

Fig. \ref{fig5}(a) presents the total number of collisions per time-step, which shows UNS is able to successfully synthesise norms at run-time to handle collisions. The results show UNS outperforms the synthesised norms set in IRON.  Furthermore, observations showed that in a lot of time-steps UNS reached zero collisions, unlike IRON. The average number of collisions in UNS is 0.08 while the average number of collisions in IRON is 0.17. 
Also, comparing the total number of collisions, the total number of collisions in UNS is 51\% lower than IRON, which shows the efficacy of the norm synthesis process.
\subsection{Scenario B (Violation Rate 70\%)}
Fig. \ref{fig4}(a) shows the average waiting time of all vehicles in UNS compared to IRON. The average  waiting time in this scenario is increased to 2.45 time-steps compared to scenario A. However, UNS still outperforms IRON, in which its average waiting time per time-step decreased from 2.75 to 2.53 time-steps. This unexplained decrease in IRON shows the essence of the primary definition of the system objectives and its incorporation  in the model. Moreover, the results show that even with a high violation rate the system objectives can be achieved using UNS.

Fig. \ref{fig4}(b) shows that UNS and IRON have quite similar range of total waiting time for priority vehicles. However, UNS outperforms IRON as the average of the total waiting time of priority vehicles is 8.92 time-steps using UNS and  9.24 time-step using IRON.

\begin{figure}%
    \centering
    \subfloat[\centering Scenario A]{{\includegraphics[width=5.66cm]{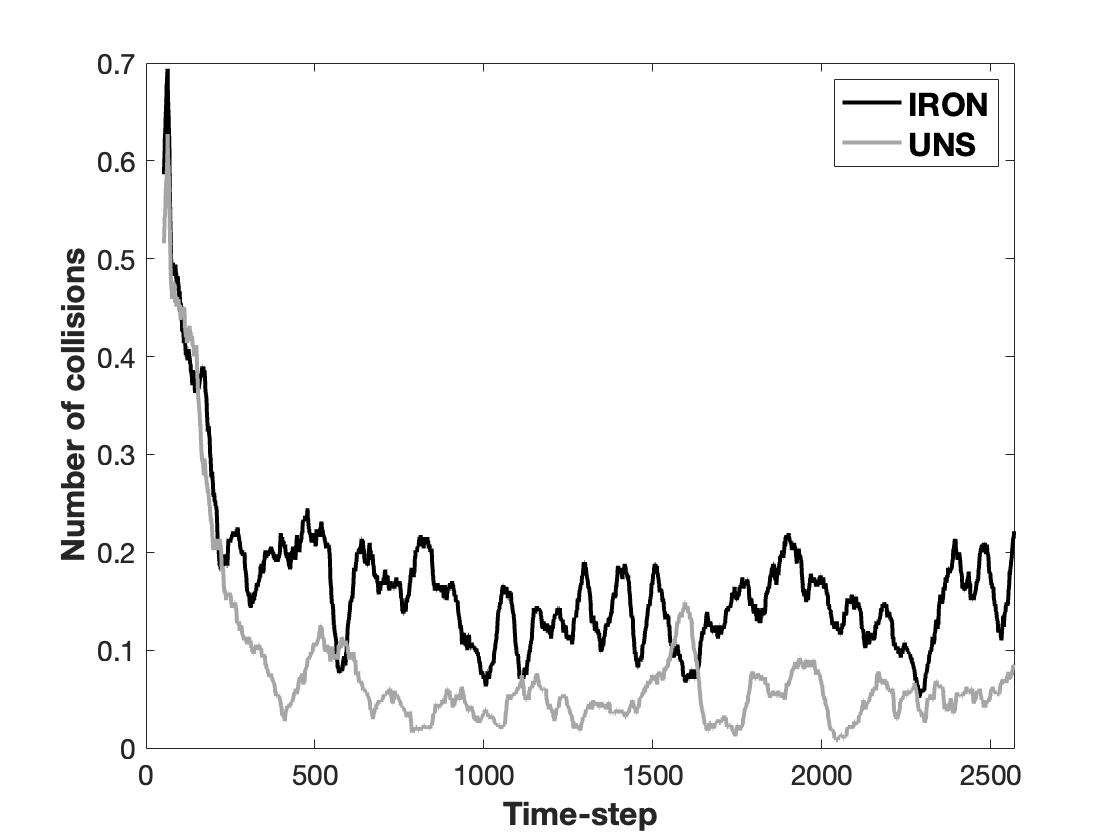}}}%
    \qquad
    \subfloat[\centering Scenario B]{{\includegraphics[width=5.66cm]{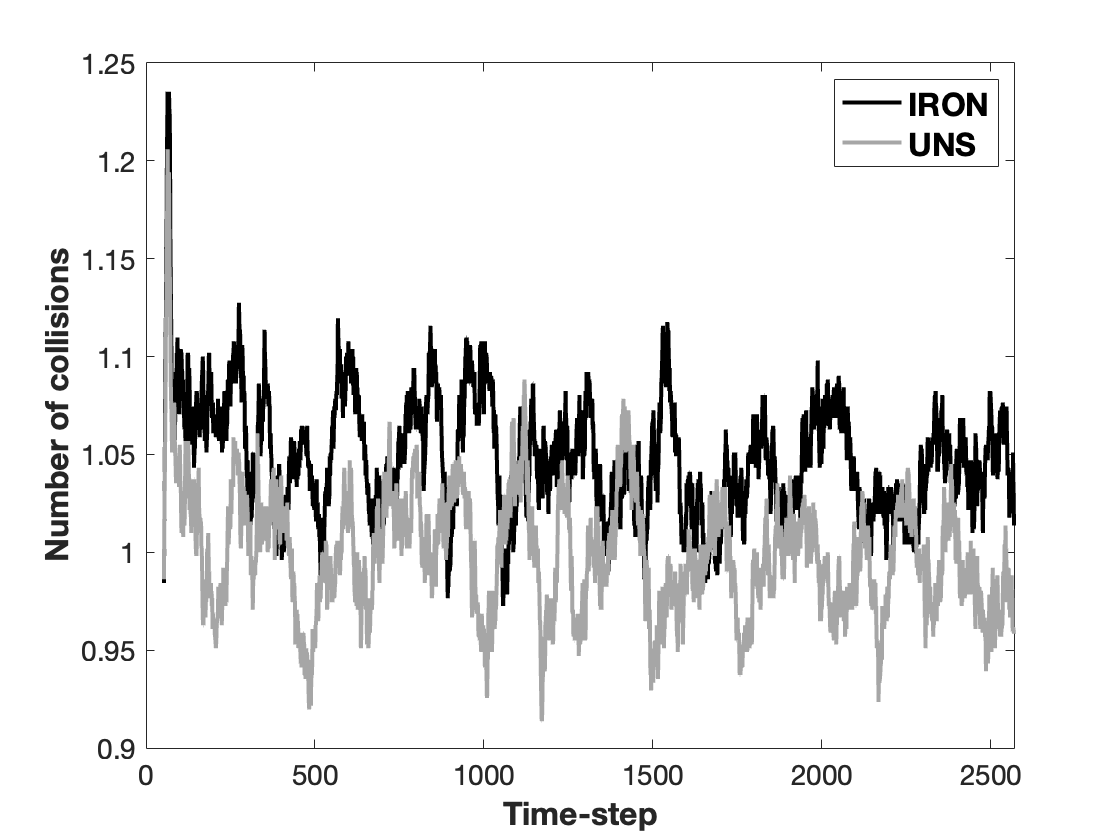}}}%
    \caption{Total number of collisions}%
    \label{fig5}%
\end{figure}

The results also show that although the violation rate has increased by 60\% compared to scenario A, the average total waiting time  of priority vehicles in UNS has only increased by 9.30\%.
Furthermore,  the number of collisions occurred  in this scenario using UNS is 4.66\% fewer compared to IRON as seen in Fig. \ref{fig5}(b).


\subsection{Scenario C (Violation Rate 0\%)}
When using 0\% violation rate with IRON simulation, IRON is not able to converge and continue the simulation. The reason behind this is that the system reaches a deadlock when all vehicles obey to the norms. Although, IRON strategy in synthesising norms relies on creating only one norm at a time, it might synthesises two unmatchable norms at different instances that when applied in the same conflict causes a deadlock. For example, if one norm is to stop if there is a vehicle on the right hand side and the second norm is to stop when there is a vehicle on the left, two lanes of the vehicles standing at the beginning of a junction will stop endlessly, when there is no violation. However, this situation does not arise in UNS because it handles unmatchable norms and if more than one norm is applicable, the utility for both norms is calculated and only one norm is applied (i.e. in the previous example, one vehicle will 'Stop' and the other will 'Go').

In all scenarios, UNS synthesises more norms than IRON. This is due to synthesising all the norms that would contribute in avoiding collision in a specific situation, supporting the idea of fairness. For example, in one of the runs IRON synthesised 15 norms, while UNS synthesised 17 norms. For example,  UNS synthesises norm $n_a$,  $n_a=(left(-),front(-),right(<),Proh(Go))$ and norm $n_b$, $n_b=(left(>),front(-),right(-),Proh(Go))$, both contributing in avoiding a collision. However, IRON only synthesises $n_b$ which will always give priority to vehicles on the right side of the intersection, and consequently  cannot support fairness.

\section{Conclusion and Future Work}
In this paper, we proposed a centralised utility-based norm synthesis (UNS) model which aims at coordinating objectives of the system with the synthesised norms in real-time. Norms in UNS are created to resolve conflicts that occur between agents and they are synthesised using case-based reasoning technique. UNS uses a utility function constructed based on the system objectives for norm reasoning. This ensures that when agents come to applying the synthesised norms, unmatchable norms and coordinated objectives of the system are handled. In addition, to ensure the effectiveness of the synthesised normative system the norms evaluation and refinement technique is inherited from IRON strategy \cite{morales2013automated}. The model was evaluated using a traffic scenario of two intersecting roads and results were compared with IRON. Results showed the efficiency of the model to meet the objectives of the system while synthesising norms in real-time. As future work, in addition to applying the model on another application domains two main directions will be followed. First, to use a decentralised architecture that involves the coordination of the agents in the process of norm synthesis. This would facilitate building several sets of norms according to each agent group's learning and objectives. Second, to transfer the norm reasoning process to be carried out in the level of agents rather than the system to ensure the agent’s autonomy in the decision-making process.
%
%
%
\bibliographystyle{splncs04}

%

\end{document}